\documentclass[manuscript=article,journal=jpcafh]{achemso}
\usepackage[dvipsnames]{xcolor}
\usepackage{subcaption}
\usepackage{amssymb}
\usepackage{amsmath}
\usepackage{braket}
\usepackage{multirow}
\usepackage{array}
\newcolumntype{L}[1]{>{\raggedright\let\newline\\\arraybackslash\hspace{0pt}}m{#1}}
\newcolumntype{C}[1]{>{\centering\let\newline\\\arraybackslash\hspace{0pt}}m{#1}}
\newcolumntype{R}[1]{>{\raggedleft\let\newline\\\arraybackslash\hspace{0pt}}m{#1}}
\usepackage[version=3]{mhchem}
\usepackage{xr}

\title{Learning Properties of Ordered and Disordered Materials from Multi-fidelity Data}

\author{Chi Chen}
\affiliation{Department of NanoEngineering, University of California San Diego, CA, USA
}

\author{Yunxing Zuo}
\affiliation{Department of NanoEngineering, University of California San Diego, CA, USA
}

\author{Weike Ye}
\affiliation{Department of NanoEngineering, University of California San Diego, CA, USA
}

\author{Xiangguo Li}
\affiliation{Department of NanoEngineering, University of California San Diego, CA, USA
}

\author{Shyue Ping Ong}%
 \email{ongsp@eng.ucsd.edu}
\affiliation{Department of NanoEngineering, University of California San Diego, CA, USA
}

\date{\today}

\begin{document}

\maketitle

\begin{abstract}
\textbf{Predicting the properties of a material from the arrangement of its atoms is a fundamental goal in materials science. While machine learning has emerged in recent years as a new paradigm to provide rapid predictions of materials properties, their practical utility is limited by the scarcity of high-fidelity data. Here, we develop multi-fidelity graph networks as a universal approach to achieve accurate predictions of materials properties with small data sizes. As a proof of concept, we show that the inclusion of low-fidelity Perdew-Burke-Ernzerhof band gaps greatly enhances the resolution of latent structural features in materials graphs, leading to a 22-45\% decrease in the mean absolute errors of experimental band gap predictions. We further demonstrate that learned elemental embeddings in materials graph networks provide a natural approach to model disorder in materials, addressing a fundamental gap in the computational prediction of materials properties.}
\end{abstract}

\textit{In silico} predictions of the properties of materials can most reliably be carried out using \textit{ab initio} calculations. However, their high computational expense and poor scalability have limited their application mostly to materials containing $< 1000$ atoms without site disorder. Further, a rule of thumb is that the more accurate the \textit{ab initio} method, the higher the computational expense and the poorer the scalability.\cite{chevrierHybridDensityFunctional2010,heydEfficientHybridDensity2004,zhangEfficientFirstprinciplesPrediction2018} It is therefore no surprise that supervised machine learning (ML) of \textit{ab initio} calculations has garnered substantial interest as a means to develop efficient surrogate models for materials property predictions.\cite{butlerMachineLearningMolecular2018} State-of-the-art ML models encode structural information (e.g., as graphs\cite{chenGraphNetworksUniversal2019,xieCrystalGraphConvolutional2018} or local environmental features\cite{behlerGeneralizedNeuralNetworkRepresentation2007,bartokGaussianApproximationPotentials2010,zuoPerformanceCostAssessment2020}) in addition to composition information, allowing them to distinguish between polymorphs that may have vastly different properties.

Most frustratingly, while building ML models from high-accuracy calculations or experiments would yield the greatest value, obtaining sufficient data to reliably train such models is also the most challenging. For example, the number of Perdew-Burke-Ernzerhof (PBE)\cite{perdewGeneralizedGradientApproximation1996} calculations in large, public databases such as the Materials Project\cite{jainCommentaryMaterialsProject2013} and Open Quantum Materials Database (OQMD)\cite{kirklinOpenQuantumMaterials2015} is on the order of $10^5-10^6$, while the number of more accurate Heyd-Scuseria-Ernzerhof (HSE)\cite{heydHybridFunctionalsBased2003} calculations is at least two orders of magnitude fewer. Similarly, while Becke, 3-parameter, Lee-Yang-Parr (B3LYP)/PBE calculations are available for millions of molecules,\cite{hachmannHarvardCleanEnergy2011} ``gold standard'' coupled-cluster single-, double-, and perturbative triple-excitations (CCSD(T)) calculations are only available for perhaps thousands of small molecules. Even fewer are the number of high-quality experimental data points.\cite{hellwegeLandoltBornsteinNumericalData1967} 

A potential approach to address this challenge is through multi-fidelity models\cite{mengCompositeNeuralNetwork2020} that combine low-fidelity data with high-fidelity data. In the handful of previous works utilizing this approach in ML of materials properties, most are two-fidelity models utilizing a co-kriging\cite{kennedyPredictingOutputComplex2000} approach, which assumes an approximately linear relationship between targets of different fidelity. The training of co-kriging models scales as $O(N^3)$ (where $N$ is the number of data points), which becomes prohibitively expensive when $N$ exceeds 10,000. Further, these efforts have been limited to specific properties of single structure prototypes.\cite{pilaniaMultifidelityMachineLearning2017,batraMultifidelityInformationFusion2019} Similarly, the transfer learning and $\Delta$-learning\cite{ramakrishnanBigDataMeets2015} are either two-fidelity approaches or non-trivial\cite{zaspelBoostingQuantumMachine2019} to extend to more than two fidelities. Multi-task neural network models\cite{dahlMultitaskNeuralNetworks2014} can handle multi-fidelity data and scale linearly with the number of data fidelities, but require homogeneous data that have all properties labeled for all the data, which is rarely the case in materials property datasets. 

\begin{figure}
\centering
\includegraphics[width=0.9\textwidth]{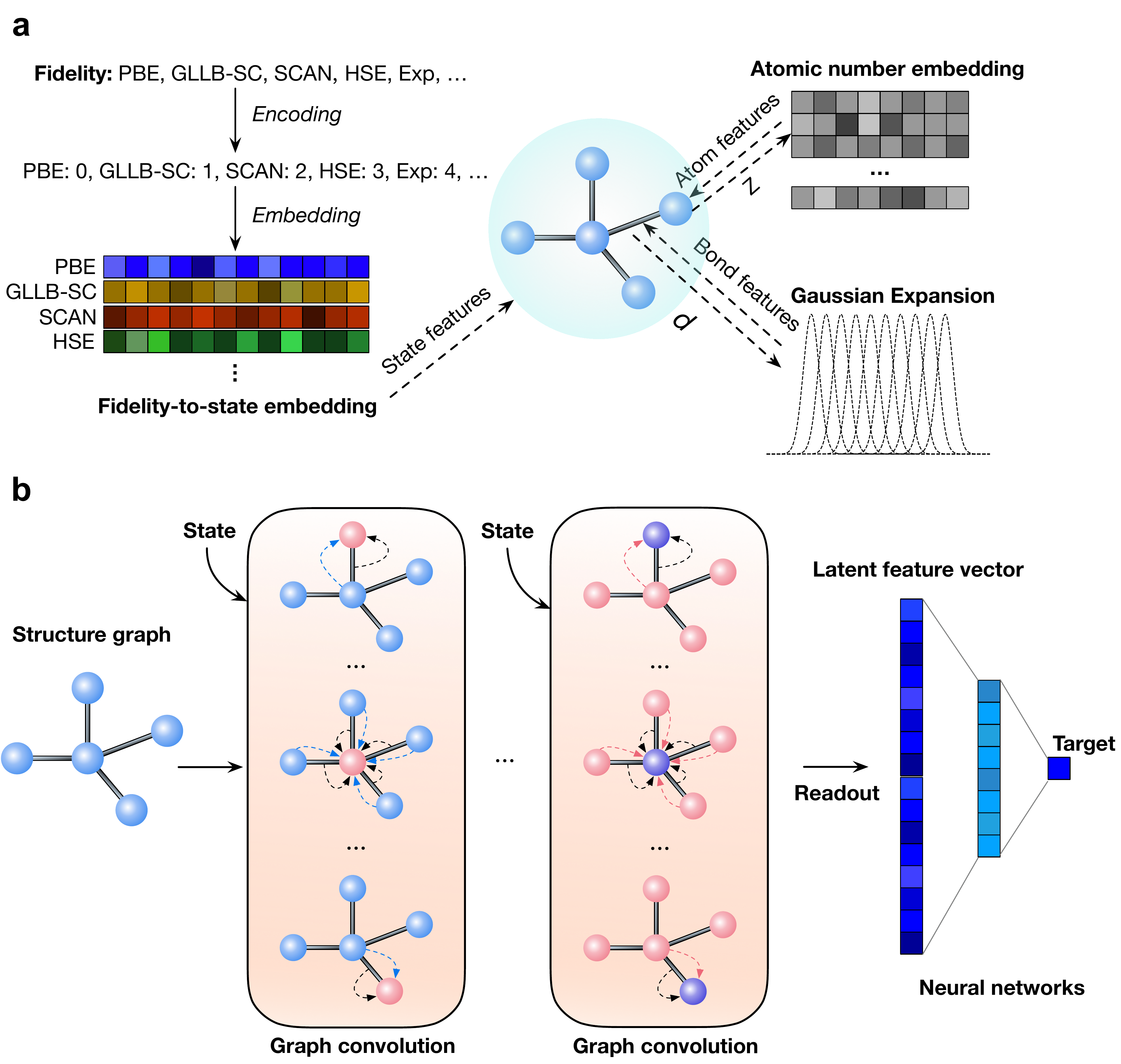}
\caption{\textbf{Multi-fidelity materials graph networks.} \textbf{a,} Representation of a material in a graph network model, with atoms as the nodes, bonds as the edges coupled with a structure-independent global state. The input atomic feature is embedded atomic number of the element. The bond feature vector is the Gaussian expanded distance. The fidelity of each data is encoded as an integer (e.g., 0 for PBE, 1 for GLLB-SC, 2 for HSE, 3 for SCAN and 4 for experiment or Exp). \textbf{b,} A materials graph network model is constructed by stacking graph convolution layers. In each graph convolution layer, sequential updates of atomic, bond and state features are performed using information from connected neighbors in the graph. The output graph in the last layer is then readout and processed in a neural network to arrive at the final prediction.}
\label{fig:schematics}
\end{figure}

Graph networks are a general, composable deep learning framework that supports both relational reasoning and combinatorial generalization.\cite{battagliaRelationalInductiveBiases2018} Previously, the current authors have shown that materials graph network (MEGNet) models significantly outperform prior ML models in predicting the properties of both molecules and crystals\cite{chenGraphNetworksUniversal2019}. However, previous graph network and other state-of-the-art models, such as the Crystal Graph Convolutional Neural Networks (CGCNN)\cite{xieCrystalGraphConvolutional2018} and SchNet\cite{schuttSchNetDeepLearning2018}, are single fidelity, typically trained on a large PBE-computed dataset, and have not been extended to multi-fidelity datasets with varying sizes. Further, all prior models have been limited to ordered materials only. Here, we develop multi-fidelity graph networks as a generalized framework for materials property prediction across computational methodologies and experiments for both ordered and disordered materials.

\section{Results}

\subsection{Multi-fidelity Graph Networks}
Fig.~\ref{fig:schematics} depicts a schematic of the multi-fidelity graph network framework. The starting point is a natural graph representation of a material, where the atoms are the nodes and the bonds between them are the edges. The input atomic attributes are simply the atomic numbers of the elements passed to a trainable elemental embedding matrix to obtain a length-16 elemental embedding vector, and the bond attributes are the Gaussian-expanded distances. The state attribute vector provides a portal for structural-independent features to be incorporated into the model. Here, the data fidelity level (e.g., computational method or experiment) is encoded as an integer and passed to a trainable fidelity embedding matrix to get a length-16  fidelity embedding vector, forming the input state attributes. A graph network model is built by applying a series of graph convolutional layers that sequentially exchange information between atoms, bonds and the state vector. In the final step, the latent features in the output graph are read out and passed into a neural network to arrive at a property prediction. Further details are available in the Methods section.

We have selected the prediction of the band gap ($E_g$) of crystals as the target problem because of its great importance in a broad range of technological applications, including photovoltaics, solar water splitting, etc., as well as the availability of data of multiple fidelities. Another application of the approach to the prediction of multi-fidelity molecular energies is demonstrated in Supplementary Section 2. The low-fidelity (low-fi) dataset comprises 52,348 PBE band gaps from the Materials Project,\cite{jainCommentaryMaterialsProject2013}. For the demonstration of model transferability, we also show the prediction of the multi-fidelity molecular energy. The high-fidelity (high-fi) computed datasets comprise 2,290 Gritsenko-Leeuwen-Lenthe-Baerends with solid correction (GLLB-SC)\cite{gritsenkoSelfconsistentApproximationKohnSham1995,kuismaKohnShamPotentialDiscontinuity2010,castelliNewLightHarvestingMaterials2015}, 472 strongly constrained and appropriately normed (SCAN)\cite{sunStronglyConstrainedAppropriately2015}\cite{borlidoLargeScaleBenchmarkExchange2019} and 6,030 Heyd-Scuseria-Ernzerhof (HSE)\cite{heydHybridFunctionalsBased2003,jieNewMaterialGoDatabase2019} band gaps. Experimental (Exp) band gaps for 2,703 ordered crystals and 278 disordered crystals\cite{zhuoPredictingBandGaps2018} are considered as a separate high-fi dataset. The computationally least expensive PBE functional is well-known to systematically underestimate the band gap,\cite{perdewPhysicalContentExact1983} and the high-fi functionals correct this to varying degrees. The data within each fidelity was randomly split into 80\% training, 10\% validation and 10\% test data and repeated six times for all models in this work. The statistics (mean and distribution) of the mean absolute errors (MAEs) on the test datasets are reported to provide an accurate assessment of model reliability.

\subsection{Band Gaps of Ordered Structures}
Single-fidelity, or 1-fi, graph network models for the band gaps of ordered crystals were first developed for each fidelity in isolation. The MAE of the 1-fi models (Fig.~\ref{fig:mf}a) is related to the data size as well as the mean absolute deviation (MAD, see Supplementary Dataset 1) within each dataset. The PBE dataset is the largest with a small MAD, and the 1-fi PBE models have the smallest average MAE of 0.27 eV. The average MAEs for the computed 1-fi models increases in the order of PBE $<$ HSE $<$ GLLB-SC $<$ SCAN, in inverse order to the dataset size. The lower MAE of the 1-fi Exp model compared to the 1-fi HSE model despite having a smaller data set size may be attributed to the relatively large fraction of metals (with zero band gap) in that dataset, which leads to smaller MAD. 

Multi-fidelity graph network models utilizing the large PBE dataset with data from other fidelities can mitigate this data quality/quantity-performance trade-off. Significantly lower average MAEs are achieved across all high-fi computed and the experimental predictions (Fig.~\ref{fig:mf}a). The 5-fi models, i.e., the models fitted using all available data, substantially improve on the MAE on the high-fi predictions over the 2-fi models, at the expense of a small increase in the MAE of the low-fi PBE predictions. The error distributions broken down in metals and non-metals for the 5-fi models are shown in Extended Data Figure 1. With the exception of the extremely small set of SCAN data on metals (17 data points), all model errors have a Gaussian-like distribution centered at zero.

Other 2-fi, 3-fi and 4-fi models, with and without PBE, were also explored (Supplementary Dataset 1). The multi-fidelity models without PBE generally have higher MAEs than the multi-fidelity models with PBE, though they typically still outperform the 1-fi models. The 4-fi models that exclude the very small SCAN dataset, i.e., models trained on PBE, GLLB-SC, HSE and Exp data, outperform the 5-fi models across all non-SCAN fidelities, which indicates that the poor quality of the SCAN dataset may have degraded performance. The reduction in average MAE of the 4-fi models over the 1-fi models range from $\sim 22\%$ for the Exp band gap to $\sim 45\%$ for the HSE band gap. Further, an increase in the number of fidelities also tends to improve model consistency, i.e., lower the standard deviation in the MAE. 

The multi-fidelity graph network models significantly outperform prior ML models in the literature. The best reported GLLB-SC model has a root-mean-squared error (RMSE) of 0.95 eV,\cite{daviesDataDrivenDiscoveryPhotoactive2019} much higher than the 4-fi RMSE of 0.68 eV on the GLLB-SC predictions. For experimental band gaps, Zhuo et al.\cite{zhuoPredictingBandGaps2018} reported MAE and RMSE of 0.75 eV and 1.46 eV for a test set of 10 compounds using a support vector regression model; the MAE and RMSE for the 4-fi model on the experimental band gap of these compounds are 0.65 eV and 1.39 eV, respectively. Zhuo et al.\cite{zhuoPredictingBandGaps2018} also reported an RMSE of 0.45 eV on the entire experimental dataset, but the dataset contains duplicated band gaps for the same composition and thus is an inaccurate metric of model performance. 

To compare alternative approaches, we have also constructed baseline 1-fi-stacked models, where a linear model is fitted for each high-fi dataset to the optimized 1-fi PBE model. This is akin to a model stacking approach and can be justified based on the relatively strong correlation between the high-fi computed and PBE band gaps (Extended Data Figure 2).\cite{morales-garciaEmpiricalPracticalWay2017} The multi-fidelity models outperform the 1-fi-stacked models, with especially large reductions in average MAEs of up to 38\% on arguably the most valuable experimental band gap predictions and 44-56\% on the GLLB-SC and HSE predictions. Another alternative approach explored is transfer learning, whereby the graph convolution layers from the PBE 1-fi models were fixed and the final output layers were retrained with higher fidelity data. Compared to the 2-fi models, the transfer learning models have somewhat lower model errors on the GLLB-SC and SCAN datasets but much higher errors on the most valuable Exp dataset, as shown in Supplementary Dataset 2. It should be noted that transfer learning is fundamentally a two-fidelity approach and requires a two-step training process. The 4-fi models outperform the transfer learning models on the Exp dataset even further. These results indicate that multi-fidelity graph network architecture is able to capture complex relationships between datasets of different fidelities. The errors of other combinations of fidelities are listed in Supplementary Dataset 1.

To gain insights into the effect of low-fi and high-fi data size on model accuracy, 2-fi models were developed using sampled subsets of each high-fi computed/experimental dataset ($N_{\mathrm{high-fi}}$) together with different quantities of data from the low-fi PBE dataset ($N_{\mathrm{PBE}}$). From Fig. \ref{fig:mf}b-e, it may be observed that adding low-fi PBE data results in a significant decrease in the average MAEs of the high-fi predictions in all cases. The average MAEs of the 2-fi models follow an approximately linear relationship with $\log_{10}{N_{\mathrm{PBE}}}$. With the exception of the SCAN 2-fi models, the magnitude of the slope decreases monotonically with an increase in $N_{\mathrm{high-fi}}$, i.e., the largest improvements are observed in the most data-constrained models. The nearly constant slope for the 2-fi SCAN models may be attributed to the extremely small size of the SCAN dataset as well as its strong correlation to the PBE dataset (Extended Data Figure 2).

\begin{figure}
\centering

\includegraphics[width=0.85\textwidth]{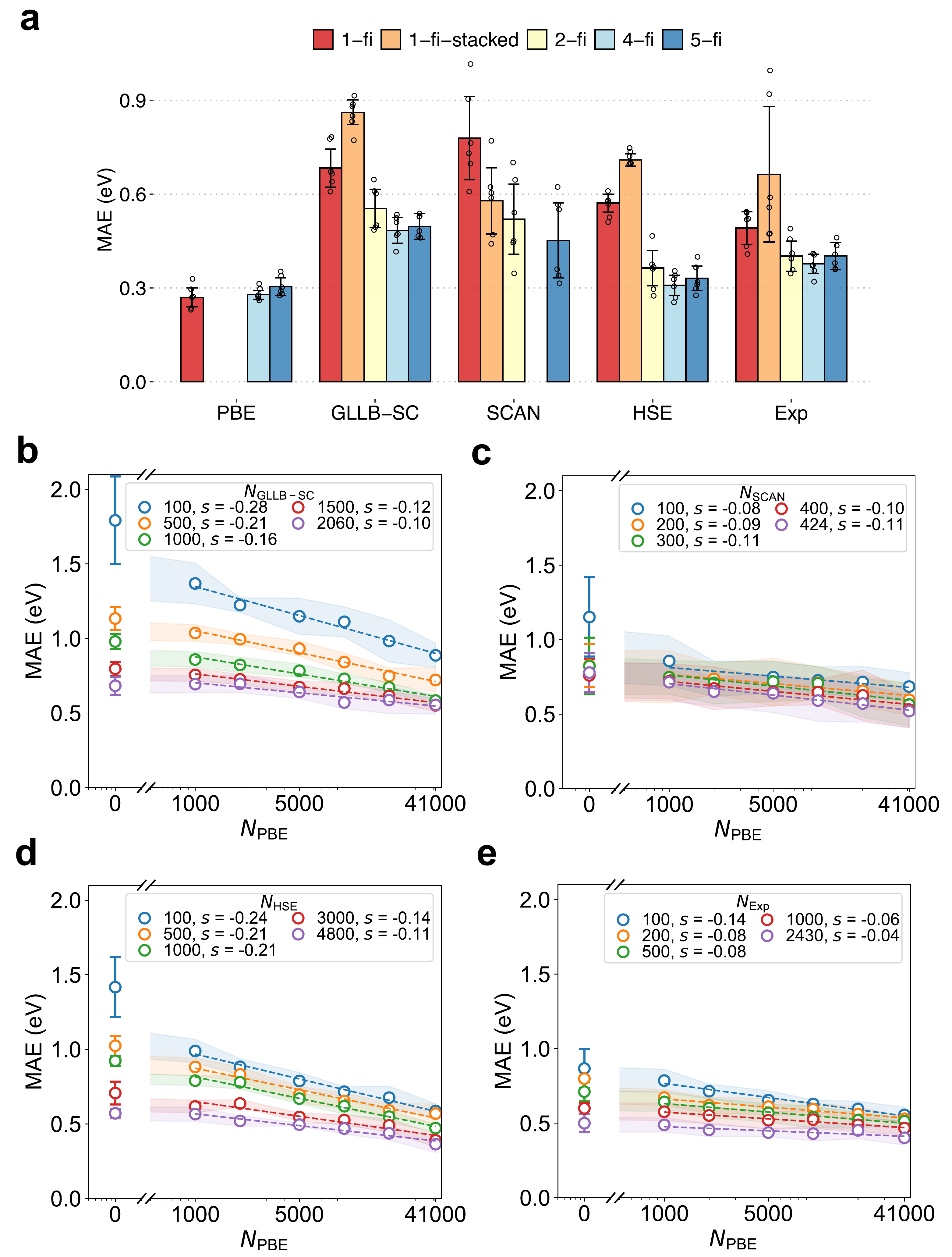}
\caption{\textbf{Test mean absolute errors (MAEs) of multi-fidelity graph network model predictions on ordered crystal band gaps.} \textbf{a,} Performance of graph network models with different fidelity combinations.  The 4-fi models used the PBE, GLLB-SC, HSE and Exp datasets, i.e., the very small SCAN dataset is excluded. All errors were obtained on the corresponding test sets of the fidelity. The error bars show one standard deviation and the dots are the individual model errors. Average MAEs of \textbf{b,} GLLB-SC, \textbf{c,} SCAN, \textbf{d,} HSE, and \textbf{e,} experimental band gaps of 2-fi models trained using sampled datasets for each high-fidelity data and PBE data. For each sub-plot, the error line is lowered with more high-fidelity data. The x-axis is plotted on a log scale and the shaded areas indicate one standard deviation of the MAE. $s$ indicates the slope for a linear fit of MAE to $\log_{10} N_{\mathrm{PBE}}$}. 
\label{fig:mf}
\end{figure}

\subsection{Latent Space Visualization}
\begin{figure}
\centering
\includegraphics[width=0.9\textwidth]{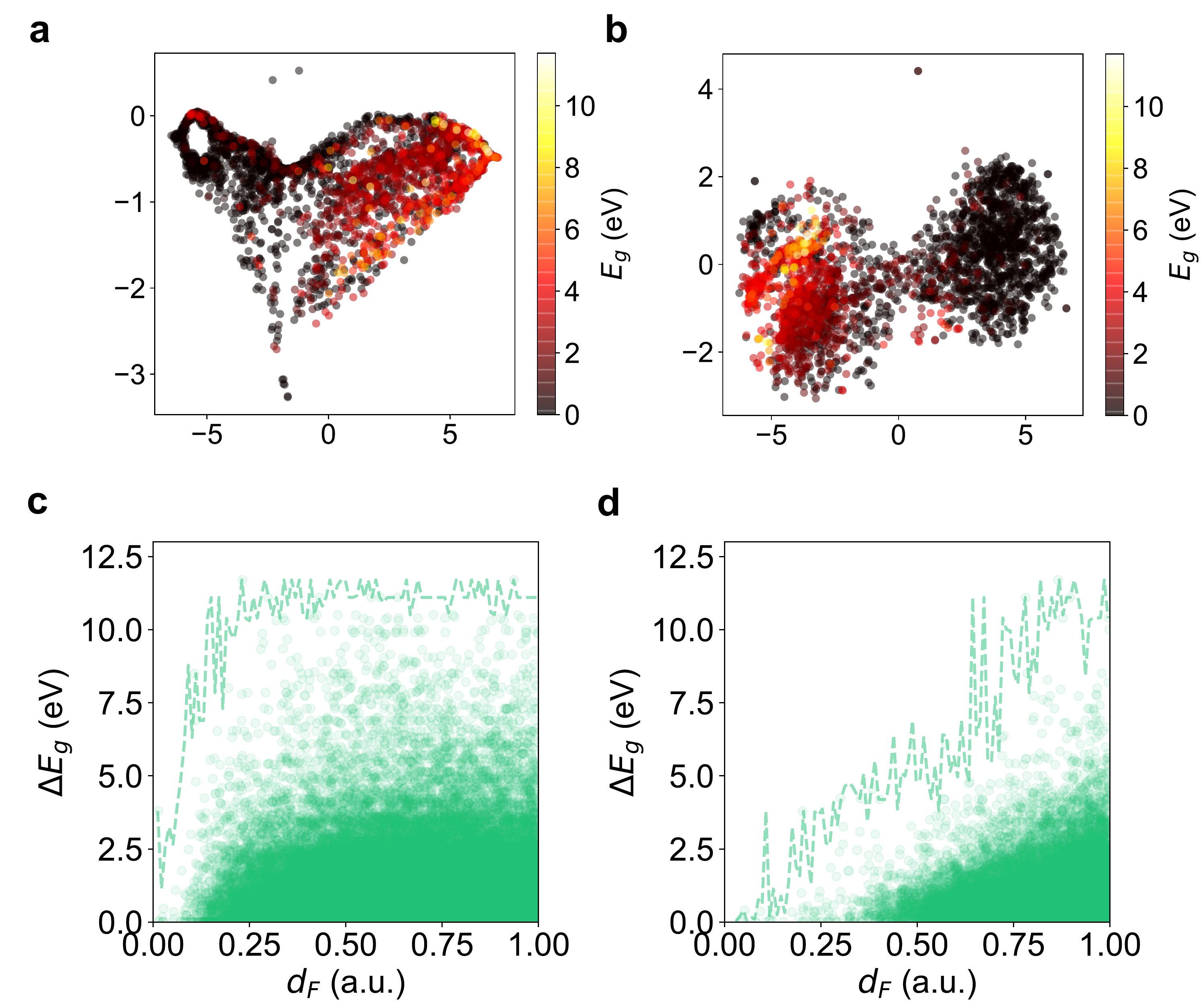}
\caption{\textbf{Effect of including low-fidelity PBE data on latent structural features.} Two dimensional t-distributed Stochastic Neighbor Embedding (complexity = 1000) projection of features for \textbf{a,} 1-fi and \textbf{b,} 2-fi models trained using 100 experimental data points and the entire PBE dataset. The markers are colored according to the experimental band gap. Plots of the experimental band gap difference ($\Delta E_g$) against normalized latent structural feature distance ($d_F$) in arbitrary units (a.u.) for the \textbf{c,} 1-fi and \textbf{d,} 2-fi-PBE models trained on all available experimental data. The dashed lines indicate the envelope of the maximum $\Delta E_g$ at each $d_F$. The scattering points are sub-sampled by a factor of 15. }
\label{fig:feature_compare}
\end{figure}

We compared the latent structural features extracted from the 1-fi and the 2-fi models trained using 100 experimental data points without and with PBE data, respectively. The t-distributed Stochastic Neighbor Embedding (t-SNE)\cite{maatenVisualizingDataUsing2008} 2D projections of the latent structure features (Fig.~\ref{fig:feature_compare}a and b) show that the inclusion of the large PBE dataset in the 2-fi model results in superior structural representations that clearly separate structures with large band gap differences.

This separation can be further quantified by plotting the band gap difference $\Delta E_g$ against the distance in the normalized structural features $d_F$ between all 3,651,753 unique pairs of crystals in the experimental data, as shown in Fig.~\ref{fig:feature_compare}c and d). The 1-fi model for experimental band gaps has poor resolution, especially for large $\Delta E_g$. A wide $d_F$ range from 0.25 to 1 corresponds to a max $\Delta E_g \sim 10$ eV, and the correspondence between $d_F$ and max $\Delta E_g$ is extremely noisy at low values. In contrast, the 2-fi model exhibits an almost linear correspondence between $d_F$ and max $\Delta E_g$ across the entire range of band gaps. Our conclusion is therefore that the inclusion of a large quantity of low-fidelity PBE data greatly assists in the learning of better latent structural features, which leads to substantially improved high-fidelity predictions. It should be noted, however, that a prerequisite for such improvements is that the low-fidelity data is of sufficient size and quality. For example, the 2-fi models without PBE perform worse than the 2-fi models with PBE (Supplementary Dataset 1).

\subsection{Band Gaps of Disordered Materials}
The multi-fidelity graph network architecture also provides a natural framework to address another major gap in the computational materials property predictions - disordered crystals. The majority of known crystals exhibit site disorder. For example, of the $\sim$ 200,000 crystals reported in the Inorganic Crystal Structure Database (ICSD),\cite{hellenbrandtInorganicCrystalStructure2004} more than 120,000 are disordered crystals. Typically, the properties of disordered crystals are estimated by sampling low energy structures among a combinatorial enumeration of distinct orderings, usually within a supercell.

In the graph network approach, we can use the learned length-16 elemental embeddings $\mathbf{W}_{Z}$ directly as the node features. In such a scheme, disordered sites can be represented as a linear combination of the elemental embeddings as $\mathbf{W}_{disordered} = \sum_{i=1} x_i\mathbf{W}_{Z_i}$, where $x_i$ is the occupancy of species $i$ in the site and $\mathbf{W}_{Z_i}$ is the element embedding for atomic number $Z_i$. Using the 4-fi model for the \textit{ordered} crystals without further retraining, the MAE of the predicted band gaps of the 278 disordered crystals in our experimental dataset is a respectable 0.63$\pm$0.14 eV, similar to the MAE of the 1-fi-stacked model on the experimental band gaps  of ordered crystals. By retraining with the disordered experimental band gap dataset, the average MAE on the experimental band gaps of disordered crystals decreases to 0.51$\pm$0.11 eV, while that of the ordered crystals remains approximately the same (0.37$\pm$0.02 eV). The average MAEs for a retrained 1-fi model on the experimental band gaps of disordered and ordered crystals are 0.55$\pm$0.13 eV and 0.50$\pm$0.07 eV, respectively. Clearly, the multi-fidelity approach improves on the performance on disordered crystals as well as ordered crystals.

\begin{figure}
\centering
\includegraphics[width=0.8\textwidth]{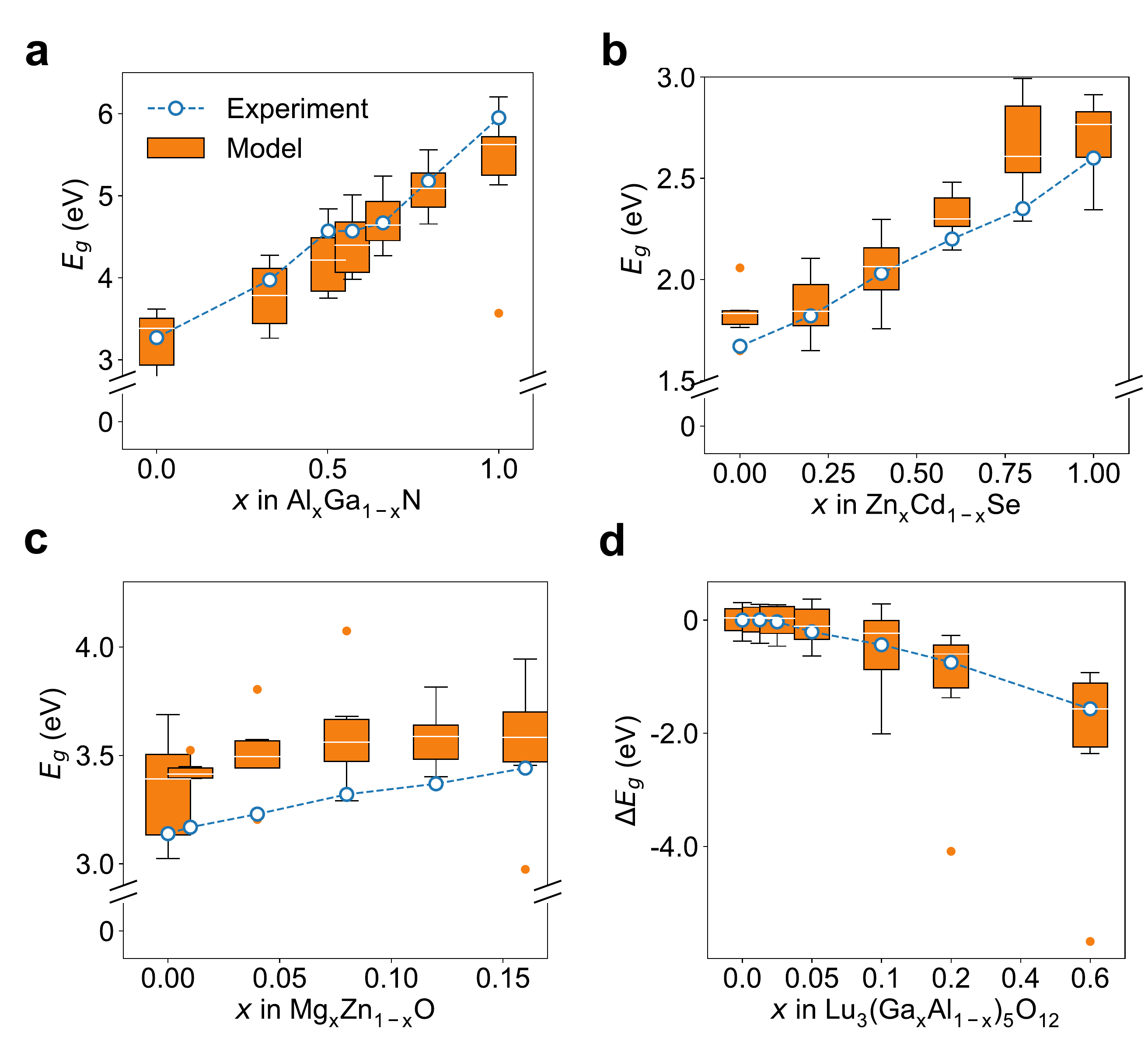}
\caption{\textbf{Performance of disordered multi-fidelity graph network models.} Predicted and experimental band gaps ($E_g$) as a function of composition variable $x$ in \textbf{a,} \ce{Al_{$x$}Ga_{$1-x$}N}, \textbf{b,} \ce{Zn_{$x$}Cd_{$1-x$}Se}, and \textbf{c,} \ce{Mg_{$x$}Zn_{$1-x$}O}. \textbf{d,} Comparison of the change in band gap with respect to \ce{Lu3Al5O12} ($\Delta E_g$) with $x$ in \ce{Lu3(Ga_{$x$}Al_{$1-x$})5O12}. The error bars indicate one standard deviation.}
\label{fig:exp_compare}
\end{figure}

To demonstrate the power of the disordered multi-fidelity graph network models, band gap engineering data was extracted from the literature for \ce{Al_{$x$}Ga_{$1-x$}N},\cite{chenBandGapEngineering2000} \ce{Zn_{$x$}Cd_{$1-x$}Se},\cite{santhoshBandGapEngineering2017}
\ce{Mg_{$x$}Zn_{$1-x$}O},\cite{ranaBandGapEngineering2015} and \ce{Lu3(Ga_{$x$}Al_{$1-x$})5O12}.\cite{fasoliBandgapEngineeringRemoving2011} The band gaps of \ce{Lu3(Ga_{$x$}Al_{$1-x$})5O12} were not present and only the band gaps of the stoichiometric endpoints for the other systems were present in our experimental dataset. The 4-fi model performs remarkably well, reproducing qualitative trends in all instances and achieving near quantitative accuracy for most systems, as shown in Fig. \ref{fig:exp_compare}. The 4-fi model reproduces the concave relationship between $x$ and the change in band gap $\Delta E_g$ for  \ce{Lu3(Ga_{$x$}Al_{$1-x$})5O12} (Fig. \ref{fig:exp_compare}d) reported by Fasoli et al.\cite{fasoliBandgapEngineeringRemoving2011}. For \ce{Zn_{$1-x$}Mg_{$x$}O}, a more pronounced concave relationship is predicted by the 4-fi model compared to the experimental measurements.\cite{ranaBandGapEngineering2015} The band gap of ZnO is notoriously poorly estimated by DFT techniques,\cite{harunDFTCalculationsElectronic2020} and even experimental measurements range from 3.1 to 3.4 eV across publications.\cite{kamarulzamanElucidationHighestValence2016} An additional proof of concept for \ce{Ba_{$y$}Sr_{$1-y$}Co_{$x$}Fe_{$1-x$}O_{$3-\delta$}} perovskite,\cite{shaoHighperformanceCathodeNext2004} a highly promising catalyst for the oxygen reduction reaction that exhibits disorder on multiple sites, is given in Extended Data Figure 3. 

\section{Discussion}

Data quality and quantity constraints are major bottlenecks to materials design. Multi-fidelity graph networks enable the efficient learning of latent structural features using large quantities of cheap low-fidelity computed data to achieve vastly improved predictions for more costly computational methods and experiments. While crystal band gaps have been selected as the model problem in this work, the multi-fidelity graph network framework is universal and readily applicable to other properties and to molecules. Two examples are provided in Extended Data Figure 4, where a large number of low-fidelity molecule calculations are shown to lead to vast improvements in the high-fidelity energy predictions.

The ability to predict band gaps of disordered materials suggests that the learned elemental embeddings in graph network models encode chemistry in a way that naturally interpolates between elements. In a traditional ML model development, each element is represented by a vector of atomic features, e.g., atomic number, electronegativity, atomic radius, etc. A disordered site, e.g., Al-Ga in  \ce{Al_{$x$}Ga_{$1-x$}N}, cannot be naturally represented as a linear combination of such feature vectors. Therefore, this is a unique attribute of our graph network models that is not present in feature-engineered ML models. The interpolation approach bears some similarity to that used in the virtual crystal approximation (VCA).\cite{nordheimElectronTheoryMetals1931} A limitation of the VCA is that it may fail in systems that do not exhibit full disorder. In the case of multi-fidelity graph networks, this limitation can be mitigated to a certain extent by applying the highly efficient models to the interpolated disordered crystal as well as ordered crystals at the same composition to arrive at a range of property predictions. Thus, multi-fidelity graph network models provide an alternative approach to \textit{in silico} materials design for the large class of disordered materials that is extremely difficult to treat with existing \textit{ab initio} computations or ML techniques.

One potential limitation of the proposed approach is the reliance on large low-fidelity datasets for learning effective latent structural representations. The low-fi dataset needs to reproduce at least general qualitative trends in the target property between different materials for such learning to be effective. For some properties, even low-fi datasets may not be available in sufficiently large quantities for effective learning. Under such instances, a transfer learning approach, i.e., where portions of a model trained on a different property is retrained on the target property, may be more appropriate.

\begin{acknowledgement}
This work was primarily supported by the Materials Project, funded by the U.S. Department of Energy, Office of Science, Office of Basic Energy Sciences, Materials Sciences and Engineering Division under contract no. DE-AC02-05-CH11231: Materials Project program KC23MP. The authors also acknowledge support from the National Science Foundation SI2-SSI Program under Award No. 1550423 for the software development portions of the work. C.C. thanks Dr. M. Horton for his assistance with the GLLB-SC data set. 
\end{acknowledgement}

\section{Author contributions} C.C. and S.P.O. conceived the idea and designed the work. C.C. implemented the models and performed the analysis. S.P.O. supervised the project.  Y.Z., W.Y. and X.L. helped with the data collection and analysis. C.C. and S.P.O. wrote the manuscript. All authors contributed to the discussion and revision.

\section{Ethics Declaration} 
\subsection{Competing Interests }
The authors declare that they have no competing financial interests.

\textbf{Correspondence} Correspondence and requests for materials should be addressed to S.P.O.~(email: ongsp@eng.ucsd.edu).

\providecommand{\latin}[1]{#1}
\makeatletter
\providecommand{\doi}
  {\begingroup\let\do\@makeother\dospecials
  \catcode`\{=1 \catcode`\}=2\doi@aux}
\providecommand{\doi@aux}[1]{\endgroup\texttt{#1}}
\makeatother
\providecommand*\mcitethebibliography{\thebibliography}
\csname @ifundefined\endcsname{endmcitethebibliography}
  {\let\endmcitethebibliography\endthebibliography}{}

\section{Figure Legends}

\begin{itemize}
    \item[] Figure 1. \textbf{Multi-fidelity materials graph networks.} \textbf{a,} Representation of a material in a graph network model, with atoms as the nodes, bonds as the edges coupled with a structure-independent global state. The input atomic feature is embedded atomic number of the element. The bond feature vector is the Gaussian expanded distance. The fidelity of each data is encoded as an integer (e.g., 0 for PBE, 1 for GLLB-SC, 2 for HSE, 3 for SCAN and 4 for experiment or Exp). \textbf{b,} A materials graph network model is constructed by stacking graph convolution layers. In each graph convolution layer, sequential updates of atomic, bond and state features are performed using information from connected neighbors in the graph. The output graph in the last layer is then readout and processed in a neural network to arrive at the final prediction.
    \item[] Figure 2. \textbf{Test mean absolute errors (MAEs) of multi-fidelity graph network model predictions on ordered crystal band gaps.} \textbf{a,} Performance of graph network models with different fidelity combinations.  The 4-fi models used the PBE, GLLB-SC, HSE and Exp datasets, i.e., the very small SCAN dataset is excluded. All errors were obtained on the corresponding test sets of the fidelity.  The error bars show one standard deviation and the dots are the individual model errors. Average MAEs of \textbf{b,} GLLB-SC, \textbf{c,} SCAN, \textbf{d,} HSE, and \textbf{e,} experimental band gaps of 2-fi models trained using sampled datasets for each high-fidelity data and PBE data. For each plot, the error line is lowered with more high-fidelity data. The x-axis is plotted on a log scale and the shaded areas indicate one standard deviation of the MAE. $s$ indicates the slope for a linear fit of MAE to $\log_{10} N_{\mathrm{PBE}}$.
    \item[] Figure 3. \textbf{Effect of including low-fidelity PBE data on latent structural features.} Two dimensional t-distributed Stochastic Neighbor Embedding (complexity = 1000) projection of features for \textbf{a,} 1-fi and \textbf{b,} 2-fi models trained using 100 experimental data points and the entire PBE dataset. The markers are colored according to the experimental band gap. Plots of the experimental band gap difference ($\Delta E_g$) against normalized latent structural feature distance ($d_F$) in arbitrary units (a.u.) for the \textbf{c,} 1-fi and \textbf{d,} 2-fi-PBE models trained on all available experimental data. The dashed lines indicate the envelope of the maximum $\Delta E_g$ at each $d_F$.
    \item[] Figure 4. \textbf{Performance of disordered multi-fidelity graph network models.} Predicted and experimental band gaps ($E_g$) as a function of composition variable $x$ in \textbf{a,} \ce{Al_{$x$}Ga_{$1-x$}N}, \textbf{b,} \ce{Zn_{$x$}Cd_{$1-x$}Se}, and \textbf{c,} \ce{Mg_{$x$}Zn_{$1-x$}O}. \textbf{d,} Comparison of the change in band gap with respect to \ce{Lu3Al5O12} ($\Delta E_g$) with $x$ in \ce{Lu3(Ga_{$x$}Al_{$1-x$})5O12}. The error bars indicate one standard deviation.
\end{itemize}

\section{Methods}

\subsection{\label{subsec:data}Data Collection and Processing}

The PBE\cite{perdewGeneralizedGradientApproximation1996} dataset comprising 52,348 crystal structures with band gaps were obtained from Materials Project\cite{jainCommentaryMaterialsProject2013} on Jun 1 2019 using the Materials Application Programming Interface in the Python Materials Genomics (pymatgen) library.\cite{ongPythonMaterialsGenomics2013,ongMaterialsApplicationProgramming2015}. The GLLB-SC band gaps from Castelli et al. \cite{castelliNewLightHarvestingMaterials2015} were obtained via MPContribs\cite{huckCommunityContributionFramework2015}. The total number of GLLB-SC band gaps is 2,290 after filtering out materials that do not have structures in the current Materials Project database and those that failed the graph computations due to abnormally long bond ($>$5 \AA). The GLLB-SC data all have positive band gaps due to the constraints applied in the structure selection in the previous work\cite{castelliNewLightHarvestingMaterials2015}. The  SCAN\cite{sunStronglyConstrainedAppropriately2015} band gaps for 472 nonmagnetic materials were obtained from Borlido et al.\cite{borlidoLargeScaleBenchmarkExchange2019}. The HSE\cite{heydHybridFunctionalsBased2003} band gaps with corresponding Materials Project structures were downloaded from the MaterialGo website.\cite{jieNewMaterialGoDatabase2019} After filtering out ill-converged calculations and those that have a much smaller HSE band gap compared to the PBE band gaps, 6,030 data points remain, of which 2,775 are metallic. Finally, the experimental band gaps were obtained from the work by Zhuo et al. \cite{zhuoPredictingBandGaps2018}. As this data set only contains compositions, the experimental crystal structure for each composition was obtained by looking up the lowest energy polymorph for a given formula in the Materials Project, followed by cross-referencing with the corresponding Inorganic Crystal Structure Database (ICSD) entry.\cite{hellenbrandtInorganicCrystalStructure2004} Further, as multiple band gap can be reported for the same composition in this data set, the band gaps for the duplicated entries were averaged. In total, 2,703 ordered (938 binary, 1306 ternary and 459 quaternary) and 278 disordered (41 binary, 132 ternary and 105 quaternary) structure-band gap pairs were obtained. All data sets are publicly available.\cite{chenLearningPropertiesOrdered2020}

\subsection{Materials Graph Networks Construction}

In materials graph networks, atoms and bonds are represented as nodes and edges in an undirected graph as $(V, E, \boldsymbol{u})$. The atom attributes $V$ are the atomic numbers $Z \in \mathbb{N}$. Each atom attribute is associated with a row vector $\mathbf{W}_{Z_i} \in \mathbb{R}^{16}$ in the element embedding matrix $\mathbf{W}_Z = [\mathbf{W}_0; \mathbf{W}_1; \mathbf{W}_2; ...; \mathbf{W}_{94}]$ where $\mathbf{W}_0$ is a dummy vector. The bond attribute is the set of Gaussian-expanded distances. For the $k$-th bond in the structure, the attributes are  
\begin{equation*}
e_{k,m} = \exp{-\frac{( d_{k} - \mu_m )^2}{\sigma^2}},   \forall d_{k} 	\leq R_c
\end{equation*}
where $d_{k}$ is the length of the bond $k$ formed by atom indices $r_k$ and $s_k$, $R_c$ is the cutoff radius and $\mu_m = \frac{m}{n_{bf}-1} \mu_{max}$ for $m = \{0, 1, 2, ..., n_{bf}-1\}$ and $n_{bf}$ is the number of bond features. In this work, $R_c= 5$\AA, $\mu_{max} = 6$ \AA, and $n_{bf} = 100$. The graphs are constructed using an edge list representation, and the edge set of the graph is represented as $E=\{(\mathbf{e}_k, r_k, s_k)\}$.  The state attributes $\mathbf{u}$ are fidelity levels $F \in \mathbb{N}$. Similar to atom attributes, fidelity $F_i$ is associated with a row vector $\mathbf{W}^f_{F_i}$ in the fidelity embedding matrix $\mathbf{W}_F = [\mathbf{W}^f_0; \mathbf{W}^f_1; \mathbf{W}^f_2; \mathbf{W}^f_3, \mathbf{W}^f_4]$. Both embedding matrices $\mathbf{W}_Z$ and $\mathbf{W}_F$ are trainable in the models, except in disordered models where the elemental embedding matrix $\mathbf{W}_Z$ is fixed to previously obtained values.

In each graph convolution layer, the graph networks are propagated sequentially as follows: 
\begin{enumerate}
    \item The attributes of each bond $k$ in the graph are updated as
\begin{equation*}
    \mathbf{e}^\prime_k = \phi_e (\mathbf{v}_{s_k}\oplus \mathbf{v}_{r_k} \oplus \mathbf{e}_k \oplus \mathbf{u})  
\end{equation*}
where $\phi_e$ is the bond update function, $\mathbf{v}_{s_k}$ and $\mathbf{v}_{r_k}$ are the atomic attributes of the two atoms forming the bond $k$, and $\oplus$ is the concatenation function.
\item Each atom $i$ is then updated as 
\begin{equation*}
    \mathbf{v}^\prime_i = \phi_v(\bar{\mathbf{v}}^e_i \oplus \mathbf{v}_i \oplus \mathbf{u})
\end{equation*}
where $\phi_v$ is the atomic update function, and $\bar{\mathbf{v}}^e_i = \operatorname*{average}_k (\mathbf{e}^\prime_k), \forall r_k = i$ is the averaged bond attributes from all bonds connected to atom $i$.
\item Finally, the state attributes are updated as
\begin{equation*}
    \mathbf{u}^\prime = \phi_u(\bar{\mathbf{u}}^e\oplus\bar{\mathbf{u}}^v\oplus\mathbf{u})
\end{equation*}
where $\phi_u$ is the state update function, and $\bar{\mathbf{u}}^e = \operatorname*{average}_k (\mathbf{e}^\prime_k)$ and $\bar{\mathbf{u}}^v = \operatorname*{average}_i (\mathbf{v}^\prime_i)$ are the averaged attributes from all atoms and bonds, respectively.
\end{enumerate}

\subsection{\label{subsec:model_training}Experimental Setup}

The models were constructed using tensorflow.\cite{abadiTensorFlowSystemLargeScale2016}. Three graph convolution layers and the \textit{Set2Set} readout function with two steps are used in the model training.\cite{chenGraphNetworksUniversal2019} The update functions in each graph convolutional layer were chosen to be multi-layer perceptron models with [64, 64, 32] hidden neurons and shifted softplus function $\ln(e^{x} + 1) - \ln{2}$ as the non-linear activation function. 

We split the data into 80\%-10\%-10\% train-validation-test ratios randomly for each fidelity independently and repeated the splitting six times. It should be noted that each structure-band gap data point is considered as a unique data point, and there are instances where the same structure with different fidelity band gaps are present in the training and validation/test data. An alternative data splitting strategy wherein structural overlaps in the training/validation/test sets are disallowed is presented in Supplementary Dataset 3. It was found that such a data splitting strategy results in significantly higher model errors, which indicates that information from multi-fidelities is necessary for the models to learn the relationships between different fidelities.

A learning rate of $\mathrm{10^{-3}}$ was used with the Adam optimizer with mean squared error as the loss function. The mean absolute error (MAE) was used as the error metric on the validation and test datasets, and the batch size for the training was set to 128. All models were trained on the corresponding training data for a maximum of 1500 epochs. During the training process, the MAE error metrics were calculated on the validation data set after each epoch. The model weights were saved after each epoch if the validation MAE reduced. Fitting was deemed to have converged if the validation MAE did not reduce for a consecutive 500 steps. An automatic training recovering mechanism was also implemented by reloading the saved model weights and reduce the learning rate by half when gradient explosion happens. For multi-fidelity model fitting, only the high-fidelity datasets without the PBE data were used in the validation set. The final model performances were evaluated on the test datasets and reported in this work.

\section{Data Availability}
Multi-fidelity band gap data and molecular data in the current study are available at \url{https://doi.org/10.6084/m9.figshare.13040330}. The data for all figures and extended data figures are available in Source Data.

\section{Code Availability}
Model fitting and results plotting codes are available at\\ \url{https://github.com/materialsvirtuallab/megnet/tree/master/multifidelity}. Materials graph network is available at
\url{https://github.com/materialsvirtuallab/megnet}. The specific version of the package can be found at \url{https://doi.org/10.5281/zenodo.4072029}.\cite{chenmegnet2020}

\renewcommand{\refname}{Methods-Only References}

\end{document}


\maketitle
\newpage

\section{Comparison with alternative models}

\begin{table}[htp]
\footnotesize
\begin{center}
\caption{\textbf{Average mean absolute errors (MAEs) of 2-fi and 4-fi graph network models trained using alternative approaches.} \rev{The first column shows the data fidelity combinations in training the models, and the other columns are the average MAEs with standard deviations on the corresponding test data fidelity.}}
\label{sitab:multi_fidelity_non_overlapping}
\begin{tabular}{  C{5.5cm} | C{1.8cm}| C{1.8cm}| C{1.8cm} | C{1.8cm} | C{1.8cm} } 
\hline
\hline
& PBE (eV) & GLLB-SC (eV) & SCAN (eV) & HSE (eV) & Exp (eV)\\
\hline
\multicolumn{6}{c}{\textbf{Non-overlapping-structure split}}\\
\hline
PBE/GLLB-SC& 0.30$\pm$0.02 & 0.57$\pm$0.02 & -& - & -\\
\hline
PBE/SCAN& 0.33$\pm$0.04 & - & 0.52$\pm$0.05 & - & -\\
\hline
PBE/HSE& 0.33$\pm$0.03 & - & - & 0.44$\pm$0.03 & -\\
\hline
PBE/Exp& 0.31$\pm$0.06 & -& - & -&0.42$\pm$0.04\\
\hline
PBE/GLLB-SC/HSE/Exp & 0.32$\pm$0.02 & 0.55$\pm$0.02  & - & 0.40$\pm$0.04 & 0.40$\pm$0.03\\
\hline
\multicolumn{6}{c}{\textbf{Positive PBE band gap}}\\
\hline
PBE/GLLB-SC & 0.35$\pm$0.02 & 0.45$\pm$0.03 & - & - & -\\
\hline
PBE/SCAN & 0.36$\pm$0.01 & - & 0.44$\pm$0.03 & - & -\\
\hline
PBE/HSE & 0.34$\pm$0.02 & - & - & 0.42$\pm$0.04 & -\\
\hline
PBE/Exp & 0.37$\pm$0.02 & - & - & - & 0.44$\pm$0.03\\
\hline
\multicolumn{6}{c}{\textbf{Inverse-data-size-weighted}}\\
\hline
PBE/GLLB-SC & 0.49$\pm$0.10 & 0.61$\pm$0.06 & - & - & - \\
\hline
PBE/SCAN & 0.78$\pm$0.32 & - & 0.65$\pm$0.15 & - & - \\
\hline
PBE/HSE & 0.38$\pm$0.07 & - & - & 0.42$\pm$0.04 & - \\
\hline
PBE/Exp & 0.52$\pm$0.03  & - & - & - & 0.48$\pm$0.05 \\
\hline
PBE/GLLB-SC/SCAN/HSE/Exp & 0.38$\pm$0.07 & 0.53$\pm$0.03 & 0.43$\pm$0.08 & 0.44$\pm$0.06 & 0.42$\pm$0.05 \\
\hline

\end{tabular}
\end{center}
\end{table}

\subsection{Non-overlapping-structure split}

The models reported in main text were trained based on a random data split procedure that treated each structure and specific fidelity band gap data point as a unique data point. We also developed 2-fi and 4-fi models where the data split is performed in such a way that a structure can only be present once regardless of fidelity in either the training, validation or test data, shown in Table \ref{sitab:multi_fidelity_non_overlapping}. In all cases, we observe a substantial increase in model errors under the non-overlapping-structure data split method, which indicates that having multi-fidelity property data for the same structure is critical \rev{for the model to learn} the relationships between different fidelities.

\subsection{Positive PBE band gap}

We also explored 2-fi models in which structures with zero PBE band gaps \rev{were excluded}. Compared to the original 2-fi models, these 2-fi models show reduced model errors for the GLLB-SC and SCAN fidelities, but increased model errors for the HSE and Exp fidelity (Table \ref{sitab:multi_fidelity_non_overlapping}). The results are consistent with the data distributions where the GLLB-SC and SCAN data are all or largely non-metals, where HSE and Exp data have considerable fraction of metallic systems.

\subsection{Inverse-data-size-weighted}

Finally, we explored the application of data weights to bias the model performance towards higher accuracy on higher fidelities. Both 2-fi and 5-fi models were developed where the weight for each data point is set as the reciprocal of the corresponding fidelity data size, i.e., the smaller size high-fidelity data sets were given much larger weights than the large PBE data. Somewhat surprisingly, these inverse-data-sized-weighted models exhibit much higher average MAEs (Table \ref{sitab:multi_fidelity_non_overlapping}) relative to corresponding uniformly weighted models. These results suggest that having high data weights on the low-fidelity is necessary to achieve good crystal representation in the deep learning models, which likely help the models on the high-fidelity tasks.

\begin{figure}[H]
\centering
\includegraphics[width=0.98\textwidth]{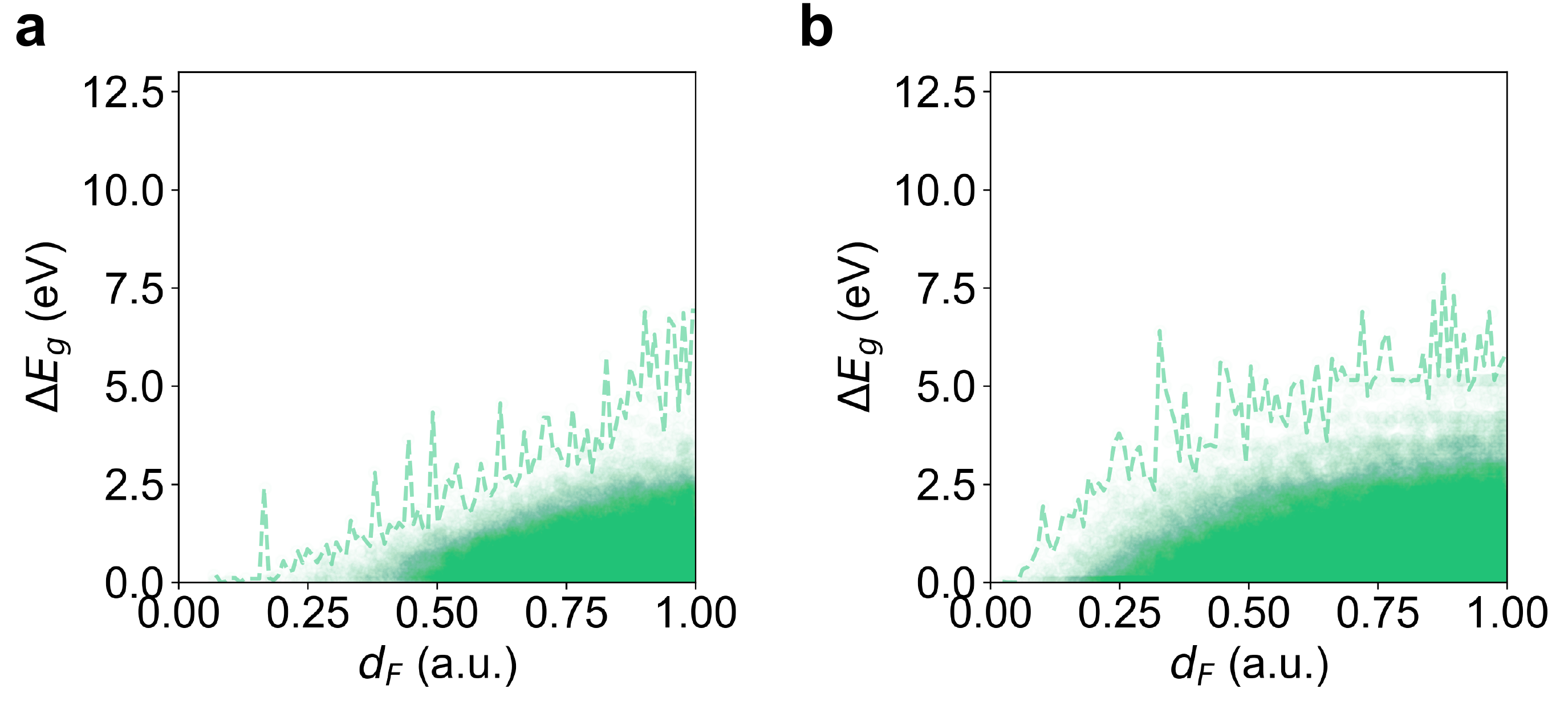}
\caption{\textbf{Material latent feature distances against the predicted experimental band gap differences for PBE/Exp 2-fi models.} Models are trained using (a) uniform data weights and (b) data weights proportional to the reciprocal of fidelity data volume.}
\label{sifig:weighted_sample_comparison}
\end{figure}

The poorer performance of the inverse-data-weighted models vis-a-vis the uniformly-weighted models is due to a decrease in the quality of the latent representations learned in these models. The PBE data set is the largest and covers a diverse range of structures and chemistries, while the other data sets are much smaller and less diverse. By decreasing the PBE weight relative to the higher fidelities, we decrease the model's ability to effectively extract effective latent features, which outweighs any potential gain in model performance by biasing the model for higher fidelities. This can be effectively demonstrated by plotting the latent feature distances between materials against the experimental band gap differences for all materials in the experimental structures for 2-fi PBE/Exp models trained on uniform weights and inverse-data-size weights (Fig. \ref{sifig:weighted_sample_comparison}). The models with uniform data weights tend to place the materials better in the latent space, as shown by the roughly linear relationship between the maximum band gap differences and the feature distances, while for the inverse-size weighted models the materials that have the feature distances $d_F$ from 0.5 to 1.0 are almost indistinguishable in terms of their band gap differences $\delta E_g$.

\subsection{Transfer learning}

We have implemented transfer learning (TL) models for all higher fidelity datasets by taking the low-fi PBE models, fixing the graph convolution layers and retraining the final layers with the higher fidelity data. The TL model MAEs are shown in Table \ref{sitab:transfer_learning}. Compared with the 2-fi models, the TL models do achieve somewhat lower MAEs on the GLLB-SC and SCAN datasets, but the errors on HSE and experimental datasets are higher. In particular, the MAE on the experimental dataset for the TL models is 0.47 eV, much higher than the 0.40 eV of the 2-fi PBE/Exp model. Further, the best performing PBE/GLLB-SC/HSE/Expt 4-fi models outperform the TL models even further.

\begin{table}[htp]
\footnotesize
\begin{center}
\caption{\textbf{Model test MAE comparisons for transfer learning, 2-fi models and 5-fi models}. The first column is the model category and the other columns are the average model MAEs with standard deviation on the corresponding test data fidelity.}
\label{sitab:transfer_learning}
\begin{tabular}{  C{3cm} | C{3cm}| C{1.8cm} | C{1.8cm} | C{1.8cm} } 
\hline
\hline
 & GLLB-SC (eV) & SCAN (eV) & HSE (eV) & Exp (eV)\\ 

\hline
\hline
Transfer learning & 0.44$\pm$0.08 & 0.35$\pm$0.04 & 0.38$\pm$0.04 & 0.47$\pm$0.04\\
\hline
2-fi & 0.55$\pm$0.06 & 0.52$\pm$0.11 & 0.36$\pm$0.06 & 0.40$\pm$0.05\\
\hline
5-fi & 0.50$\pm$0.04 & 0.45$\pm$0.12 & 0.33$\pm$0.04 & 0.40$\pm$0.04\\
\hline
\hline
\end{tabular}
\end{center}
\end{table}

\begin{table}[htp]
\footnotesize
\begin{center}
\caption{\textbf{The GLLB-SC test MAEs for two-fidelity PBE/GLLB-SC models with various low-fidelity and high-fidelity data size combinations.} }
\label{sitab:two_fide_details_gllb}
\begin{tabular}{  C{1.8cm}| C{1.8cm}| C{3cm} | C{3cm} | C{3cm} } 
\hline
\hline
 High-fi size & Low-fi size & MAE mean (eV) & MAE std (eV) & MAE range (eV)\\
\hline
\hline
100 & 0 & 1.79 & 0.29 & [1.39, 2.29] \\ 
\hline 
100 & 1000 & 1.37 & 0.14 & [1.14, 1.57] \\ 
\hline 
100 & 2000 & 1.22 & 0.05 & [1.15, 1.29] \\ 
\hline 
100 & 5000 & 1.15 & 0.12 & [0.99, 1.31] \\ 
\hline 
100 & 10000 & 1.11 & 0.10 & [0.93, 1.25] \\ 
\hline 
100 & 20000 & 0.98 & 0.14 & [0.80, 1.16] \\ 
\hline 
100 & 41000 & 0.89 & 0.08 & [0.74, 1.00] \\ 
\hline 
500 & 0 & 1.13 & 0.08 & [1.02, 1.22] \\ 
\hline 
500 & 1000 & 1.04 & 0.06 & [0.94, 1.10] \\ 
\hline 
500 & 2000 & 0.99 & 0.04 & [0.93, 1.06] \\ 
\hline 
500 & 5000 & 0.93 & 0.05 & [0.87, 1.02] \\ 
\hline 
500 & 10000 & 0.84 & 0.06 & [0.75, 0.94] \\ 
\hline 
500 & 20000 & 0.75 & 0.05 & [0.67, 0.82] \\ 
\hline 
500 & 41000 & 0.72 & 0.08 & [0.60, 0.86] \\ 
\hline 
1000 & 0 & 0.98 & 0.05 & [0.91, 1.06] \\ 
\hline 
1000 & 1000 & 0.86 & 0.05 & [0.78, 0.94] \\ 
\hline 
1000 & 2000 & 0.82 & 0.05 & [0.78, 0.93] \\ 
\hline 
1000 & 5000 & 0.78 & 0.05 & [0.72, 0.84] \\ 
\hline 
1000 & 10000 & 0.73 & 0.03 & [0.67, 0.77] \\ 
\hline 
1000 & 20000 & 0.67 & 0.03 & [0.64, 0.73] \\ 
\hline 
1000 & 41000 & 0.58 & 0.03 & [0.53, 0.64] \\ 
\hline 
1500 & 0 & 0.80 & 0.05 & [0.72, 0.88] \\ 
\hline 
1500 & 1000 & 0.76 & 0.04 & [0.68, 0.82] \\ 
\hline 
1500 & 2000 & 0.73 & 0.03 & [0.68, 0.78] \\ 
\hline 
1500 & 5000 & 0.67 & 0.04 & [0.61, 0.73] \\ 
\hline 
1500 & 10000 & 0.66 & 0.04 & [0.59, 0.72] \\ 
\hline 
1500 & 20000 & 0.62 & 0.03 & [0.58, 0.67] \\ 
\hline 
1500 & 41000 & 0.55 & 0.04 & [0.50, 0.60] \\ 
\hline 
2060 & 0 & 0.68 & 0.06 & [0.60, 0.76] \\ 
\hline 
2060 & 1000 & 0.69 & 0.06 & [0.61, 0.79] \\ 
\hline 
2060 & 2000 & 0.70 & 0.06 & [0.59, 0.77] \\ 
\hline 
2060 & 5000 & 0.64 & 0.04 & [0.60, 0.73] \\ 
\hline 
2060 & 10000 & 0.57 & 0.04 & [0.50, 0.62] \\ 
\hline 
2060 & 20000 & 0.59 & 0.09 & [0.50, 0.75] \\ 
\hline 
2060 & 41000 & 0.55 & 0.06 & [0.48, 0.65] \\ 
\hline 
\end{tabular}
\end{center}
\end{table}

\begin{table}[htp]
\footnotesize
\begin{center}
\caption{\textbf{The SCAN test MAEs for two-fidelity PBE/SCAN models with various low-fidelity and high-fidelity data size combinations.} }
\label{sitab:two_fide_details_scan}
\begin{tabular}{  C{1.8cm}| C{1.8cm}| C{3cm} | C{3cm} | C{3cm} } 
\hline
\hline
 High-fi size & Low-fi size & MAE mean (eV) & MAE std (eV) & MAE range (eV)\\
\hline
\hline
100 & 0 & 1.15 & 0.27 & [0.81, 1.68] \\ 
\hline 
100 & 1000 & 0.86 & 0.17 & [0.56, 1.09] \\ 
\hline 
100 & 2000 & 0.74 & 0.12 & [0.57, 0.92] \\ 
\hline 
100 & 5000 & 0.75 & 0.11 & [0.64, 0.92] \\ 
\hline 
100 & 10000 & 0.73 & 0.09 & [0.57, 0.86] \\ 
\hline 
100 & 20000 & 0.72 & 0.10 & [0.55, 0.83] \\ 
\hline 
100 & 41000 & 0.68 & 0.09 & [0.59, 0.85] \\ 
\hline 
200 & 0 & 0.83 & 0.14 & [0.66, 1.10] \\ 
\hline 
200 & 1000 & 0.75 & 0.18 & [0.54, 1.01] \\ 
\hline 
200 & 2000 & 0.74 & 0.10 & [0.64, 0.89] \\ 
\hline 
200 & 5000 & 0.71 & 0.16 & [0.55, 1.00] \\ 
\hline 
200 & 10000 & 0.71 & 0.11 & [0.57, 0.88] \\ 
\hline 
200 & 20000 & 0.66 & 0.10 & [0.54, 0.80] \\ 
\hline 
200 & 41000 & 0.60 & 0.11 & [0.46, 0.78] \\ 
\hline 
300 & 0 & 0.82 & 0.19 & [0.61, 1.19] \\ 
\hline 
300 & 1000 & 0.75 & 0.11 & [0.57, 0.92] \\ 
\hline 
300 & 2000 & 0.71 & 0.14 & [0.54, 0.90] \\ 
\hline 
300 & 5000 & 0.72 & 0.13 & [0.55, 0.94] \\ 
\hline 
300 & 10000 & 0.71 & 0.11 & [0.57, 0.87] \\ 
\hline 
300 & 20000 & 0.61 & 0.13 & [0.45, 0.82] \\ 
\hline 
300 & 41000 & 0.57 & 0.14 & [0.39, 0.75] \\ 
\hline 
400 & 0 & 0.76 & 0.11 & [0.61, 0.95] \\ 
\hline 
400 & 1000 & 0.72 & 0.12 & [0.54, 0.94] \\ 
\hline 
400 & 2000 & 0.67 & 0.14 & [0.50, 0.92] \\ 
\hline 
400 & 5000 & 0.65 & 0.09 & [0.51, 0.81] \\ 
\hline 
400 & 10000 & 0.65 & 0.09 & [0.54, 0.82] \\ 
\hline 
400 & 20000 & 0.63 & 0.17 & [0.49, 0.90] \\ 
\hline 
400 & 41000 & 0.53 & 0.12 & [0.34, 0.70] \\ 
\hline 
424 & 0 & 0.78 & 0.13 & [0.61, 1.01] \\ 
\hline 
424 & 1000 & 0.71 & 0.13 & [0.59, 0.89] \\ 
\hline 
424 & 2000 & 0.65 & 0.10 & [0.56, 0.86] \\ 
\hline 
424 & 5000 & 0.64 & 0.13 & [0.53, 0.84] \\ 
\hline 
424 & 10000 & 0.59 & 0.14 & [0.43, 0.80] \\ 
\hline 
424 & 20000 & 0.57 & 0.12 & [0.38, 0.74] \\ 
\hline 
424 & 41000 & 0.52 & 0.11 & [0.35, 0.68] \\ 
\hline 
\hline 
\end{tabular}
\end{center}
\end{table}

\begin{table}[htp]
\footnotesize
\begin{center}
\caption{\textbf{The HSE test MAEs for two-fidelity PBE/HSE models with various low-fidelity and high-fidelity data size combinations.} }
\label{sitab:two_fide_details_hse}
\begin{tabular}{  C{1.8cm}| C{1.8cm}| C{3cm} | C{3cm} | C{3cm} } 
\hline
\hline
 High-fi size & Low-fi size & MAE mean (eV) & MAE std (eV) & MAE range (eV)\\
\hline
\hline
100 & 0 & 1.42 & 0.20 & [1.22, 1.76] \\ 
\hline 
100 & 1000 & 0.99 & 0.08 & [0.89, 1.09] \\ 
\hline 
100 & 2000 & 0.88 & 0.05 & [0.80, 0.95] \\ 
\hline 
100 & 5000 & 0.79 & 0.06 & [0.72, 0.87] \\ 
\hline 
100 & 10000 & 0.72 & 0.03 & [0.69, 0.77] \\ 
\hline 
100 & 20000 & 0.68 & 0.08 & [0.57, 0.76] \\ 
\hline 
100 & 41000 & 0.59 & 0.05 & [0.53, 0.65] \\ 
\hline 
500 & 0 & 1.02 & 0.07 & [0.92, 1.12] \\ 
\hline 
500 & 1000 & 0.88 & 0.06 & [0.82, 0.97] \\ 
\hline 
500 & 2000 & 0.83 & 0.05 & [0.76, 0.90] \\ 
\hline 
500 & 5000 & 0.70 & 0.05 & [0.64, 0.80] \\ 
\hline 
500 & 10000 & 0.65 & 0.06 & [0.56, 0.71] \\ 
\hline 
500 & 20000 & 0.59 & 0.05 & [0.51, 0.64] \\ 
\hline 
500 & 41000 & 0.57 & 0.07 & [0.48, 0.65] \\ 
\hline 
1000 & 0 & 0.92 & 0.03 & [0.88, 0.97] \\ 
\hline 
1000 & 1000 & 0.79 & 0.03 & [0.75, 0.85] \\ 
\hline 
1000 & 2000 & 0.78 & 0.05 & [0.70, 0.83] \\ 
\hline 
1000 & 5000 & 0.67 & 0.03 & [0.64, 0.73] \\ 
\hline 
1000 & 10000 & 0.62 & 0.05 & [0.56, 0.71] \\ 
\hline 
1000 & 20000 & 0.55 & 0.04 & [0.49, 0.60] \\ 
\hline 
1000 & 41000 & 0.47 & 0.03 & [0.45, 0.52] \\ 
\hline 
3000 & 0 & 0.71 & 0.08 & [0.59, 0.81] \\ 
\hline 
3000 & 1000 & 0.62 & 0.04 & [0.55, 0.66] \\ 
\hline 
3000 & 2000 & 0.64 & 0.03 & [0.60, 0.67] \\ 
\hline 
3000 & 5000 & 0.55 & 0.03 & [0.52, 0.59] \\ 
\hline 
3000 & 10000 & 0.53 & 0.03 & [0.47, 0.57] \\ 
\hline 
3000 & 20000 & 0.49 & 0.04 & [0.42, 0.56] \\ 
\hline 
3000 & 41000 & 0.39 & 0.05 & [0.30, 0.43] \\ 
\hline 
4800 & 0 & 0.57 & 0.03 & [0.53, 0.61] \\ 
\hline 
4800 & 1000 & 0.57 & 0.05 & [0.49, 0.64] \\ 
\hline 
4800 & 2000 & 0.52 & 0.04 & [0.46, 0.55] \\ 
\hline 
4800 & 5000 & 0.49 & 0.01 & [0.48, 0.51] \\ 
\hline 
4800 & 10000 & 0.47 & 0.04 & [0.43, 0.54] \\ 
\hline 
4800 & 20000 & 0.44 & 0.04 & [0.40, 0.53] \\ 
\hline 
4800 & 41000 & 0.36 & 0.06 & [0.29, 0.46] \\ 
\hline 
\hline 
\end{tabular}
\end{center}
\end{table}

\begin{table}[htp]
\footnotesize
\begin{center}
\caption{\textbf{The Exp test MAEs for two-fidelity PBE/Exp models with various low-fidelity and high-fidelity data size combinations.} }
\label{sitab:two_fide_details_exp}
\begin{tabular}{  C{1.8cm}| C{1.8cm}| C{3cm} | C{3cm} | C{3cm} } 
\hline
\hline
 High-fi size & Low-fi size & MAE mean (eV) & MAE std (eV) & MAE range (eV)\\
\hline
\hline
100 & 0 & 0.87 & 0.13 & [0.75, 1.08] \\ 
\hline 
100 & 1000 & 0.79 & 0.08 & [0.68, 0.93] \\ 
\hline 
100 & 2000 & 0.71 & 0.05 & [0.63, 0.79] \\ 
\hline 
100 & 5000 & 0.65 & 0.06 & [0.54, 0.73] \\ 
\hline 
100 & 10000 & 0.63 & 0.03 & [0.58, 0.67] \\ 
\hline 
100 & 20000 & 0.60 & 0.05 & [0.52, 0.69] \\ 
\hline 
100 & 41000 & 0.56 & 0.05 & [0.48, 0.64] \\ 
\hline 
200 & 0 & 0.80 & 0.08 & [0.64, 0.88] \\ 
\hline 
200 & 1000 & 0.67 & 0.05 & [0.59, 0.76] \\ 
\hline 
200 & 2000 & 0.62 & 0.05 & [0.55, 0.71] \\ 
\hline 
200 & 5000 & 0.61 & 0.05 & [0.54, 0.70] \\ 
\hline 
200 & 10000 & 0.60 & 0.04 & [0.55, 0.68] \\ 
\hline 
200 & 20000 & 0.56 & 0.05 & [0.48, 0.61] \\ 
\hline 
200 & 41000 & 0.53 & 0.02 & [0.50, 0.55] \\ 
\hline 
500 & 0 & 0.71 & 0.06 & [0.59, 0.79] \\ 
\hline 
500 & 1000 & 0.64 & 0.07 & [0.52, 0.74] \\ 
\hline 
500 & 2000 & 0.60 & 0.07 & [0.49, 0.70] \\ 
\hline 
500 & 5000 & 0.57 & 0.06 & [0.50, 0.69] \\ 
\hline 
500 & 10000 & 0.54 & 0.07 & [0.44, 0.66] \\ 
\hline 
500 & 20000 & 0.52 & 0.05 & [0.44, 0.59] \\ 
\hline 
500 & 41000 & 0.51 & 0.02 & [0.47, 0.54] \\ 
\hline 
1000 & 0 & 0.60 & 0.04 & [0.55, 0.66] \\ 
\hline 
1000 & 1000 & 0.58 & 0.06 & [0.47, 0.66] \\ 
\hline 
1000 & 2000 & 0.55 & 0.06 & [0.45, 0.63] \\ 
\hline 
1000 & 5000 & 0.52 & 0.06 & [0.42, 0.60] \\ 
\hline 
1000 & 10000 & 0.52 & 0.06 & [0.44, 0.61] \\ 
\hline 
1000 & 20000 & 0.49 & 0.05 & [0.43, 0.57] \\ 
\hline 
1000 & 41000 & 0.47 & 0.05 & [0.39, 0.51] \\ 
\hline 
2430 & 0 & 0.50 & 0.06 & [0.42, 0.60] \\ 
\hline 
2430 & 1000 & 0.49 & 0.05 & [0.41, 0.56] \\ 
\hline 
2430 & 2000 & 0.45 & 0.05 & [0.38, 0.52] \\ 
\hline 
2430 & 5000 & 0.44 & 0.03 & [0.40, 0.48] \\ 
\hline 
2430 & 10000 & 0.43 & 0.05 & [0.36, 0.51] \\ 
\hline 
2430 & 20000 & 0.45 & 0.06 & [0.37, 0.54] \\ 
\hline 
2430 & 41000 & 0.40 & 0.05 & [0.33, 0.48] \\ 
\hline 
\hline 
\hline 
\end{tabular}
\end{center}
\end{table}

\clearpage

\section{Multi-fidelity models for energies of molecules \label{sec:molecules}}

To establish the universal applicability of the multi-fidelity graph network framework to molecules and other properties beyond band gaps, models were developed using computed molecule energies from the QM9\cite{ramakrishnanQuantumChemistryStructures2014} dataset and the QM7b dataset.\cite{zaspelBoostingQuantumMachine2019} The procedures for training (data split, hyperparameter optimization, etc.) are similar to those for the crystal band gap models, as outlined in the Methods section of the main manuscript.

The QM9 dataset contains B3LYP/6-31G(2df,p) calculations of 130,462 small molecules consisting of C, H, O, N, F elements. For a subset of 6,095 \ce{C7H10O2}, the energies, enthalpies and free energies are also calculated using more accurate/expensive G4MP2 level of theory. Extended Data Fig. 4a shows the effect adding the low-fidelity B3LYP 
data on the prediction average mean absolute error (MAE) of high-fidelity G4MP2 energies. With the addition of the B3LYP data, the 2-fi models reaches chemical accuracy of 1 kcal/mol on the average MAE of the G4MP2 energy predictions with only $\sim$300 G4MP2 data points. Without B3LYP data, i.e., a 1-fi model, chemical accuracy can only be achieved with $\sim$ 4000 G4MP2 data points. In other words, the 2-fi models converges to within chemical accuracy with an order of magnitude fewer computationally costly G4MP2 data points.

The QM7b dataset contains HF, MP2, and CCSD(T) energy calculations with the cc-pvdz basis set of 7,211 molecules of C, H, O, N, F, Cl, S elements.\cite{zaspelBoostingQuantumMachine2019} Extended Data Fig. 4b shows the effect of adding low-fidelity HF and MP2 data on average MAEs of the the multi-fidelity models on the CCSD(T) energy predictions. In all cases, the multi-fidelity models significantly outperform the 1-fi CCSD(T) models, achieving lower MAEs on the high-fidelity CCSD(T) energies with much fewer CCSD(T) data points. A larger data size ratio $s$ (greater quantity of low-fidelity HF and MP2 calculations) leads to a more rapid decrease in MAE with the addition of CCSD(T) data. Most interestingly, the inclusion of a large quantity of computationally cheap HF data in a 2-fi HF/CCSD(T) model (gray lines in Extended Data Fig. 4b) achieves essentially the same improvements in MAE as a 3-fi HF/MP2/CCSD(T) model (blue lines in Extended Data Fig. 4b) that requires more computationally expensive MP2 calculations. Close to chemical accuracy of 1 kcal/mol can be achieved on the average MAE with about O($10^2$) costly CCSD(T) datapoints when $16\times$ the quantity of HF data is included in a 2-fi model.
\clearpage

\providecommand{\latin}[1]{#1}
\makeatletter
\providecommand{\doi}
  {\begingroup\let\do\@makeother\dospecials
  \catcode`\{=1 \catcode`\}=2\doi@aux}
\providecommand{\doi@aux}[1]{\endgroup\texttt{#1}}
\makeatother
\providecommand*\mcitethebibliography{\thebibliography}
\csname @ifundefined\endcsname{endmcitethebibliography}
  {\let\endmcitethebibliography\endthebibliography}{}